# EXACT ZERO-ENERGY SOLUTION

# FOR A NEW FAMILY OF ANHARMONIC POTENTIALS

Yoël Lana-Renault

*Departamento de Física Teórica*

*Universidad de Zaragoza. 50009 Zaragoza. Spain*

e-mail: *yoel@kepler.unizar.es*

**Abstract.** An explicit, oscillatory solution of the type $r = [r_o + A_o \cdot \sin(\omega t + \phi_o)]^a$ is found for the zero-energy one-dimensional motion of a particle under a specific family of anharmonic potentials.

## I. The Potential

In classical mechanics there are very few non-trivial potentials that admit an explicit analytical solution. Our aim in this paper is to present a new family of one-dimensional anharmonic potentials represented by the function:

$$V(r) = \frac{1}{2} \cdot m \cdot a^2 \cdot \omega^2 \cdot r^2 \cdot \left[ 1 - 2 \cdot r_o \cdot r^{-\frac{1}{a}} + \left( r_o^2 - A_o^2 \right) \cdot r^{-\frac{2}{a}} \right] \qquad (1)$$

for which one is able to find an analytical solution.

In (1), r is the position of a point-like particle of mass m; $r_o$ and $A_o$ are two independent parameters, fulfilling $r_o > A_o$. As usual, $\omega$ will represent the angular frequency of the oscillatory solutions; -a- is a free non-null parameter, and we will study the equation of the oscillation of zero-energy around the minimum of the potential.

For a large enough r, V(r) is always confining. If $a < 0$, the third term within the brackets is the leader. If $a > 0$, then, the first term in the bracket produces a -parabolic- confinement. Our study will be devoted to the analysis of the oscillations generated around the potential minimum, which has at its right the confining wall just mentioned.

At the origin, $r = 0$, V(r) can exhibit different behaviours, but they are of no interest here.

Beyond the origin, V(r) has, in general, a minimum, at $r = R_1$, and a maximum, at $r = R_2$. These values are

$$R_1 = \left[ r_o + \frac{-r_o + \sqrt{r_o^2 + 4 \cdot a \cdot (a-1) \cdot A_o^2}}{2 \cdot a} \right]^a \quad (2)$$

$$R_2 = \left[ r_o + \frac{-r_o - \sqrt{r_o^2 + 4 \cdot a \cdot (a-1) \cdot A_o^2}}{2 \cdot a} \right]^a \quad (3)$$

Note that $R_1$ always exists. $R_2$, on the contrary, is not real for $0 < a < 1$.

The oscillatory motion is confined to the region where $V(r) < E$. Thus, for oscillations of $E = 0$, imposing the condition $V(r) = 0$ we find:

If $a > 0$, $R_M = (r_o + A_o)^a$ and $R_m = (r_o - A_o)^a$ (4)

If $a < 0$, $R_M = (r_o - A_o)^a$ and $R_m = (r_o + A_o)^a$ (5)

where $R_M$ and $R_m$ are the maximum and minimum limiting distances, respectively. In all cases, $R_2 < R_m < R_1 < R_M$.

## II. Several Examples

In the five cases depicted in Sections II and III, we will assume that $m = 1$, $\omega = 10$, $r_o = 2$ and $A_o = 1$.

The purpose of these drawings is to give an idea of the several types of geometries included in Eq. (1). In all cases the continuous line represents the potential, whilst the dotted straight line represents the $E = 0$ level.

Let us consider first $a = -1.1$ $(a < 0)$. This is illustrated in Fig. 1. Some of the results are:

$$R_2 = 0.188 \quad , \quad R_m = 0.299 \quad , \quad R_1 = 0.779 \quad , \quad R_M = 1$$

Here $V(R_2) = 0.574$ and $V(R_1) = -10.37$

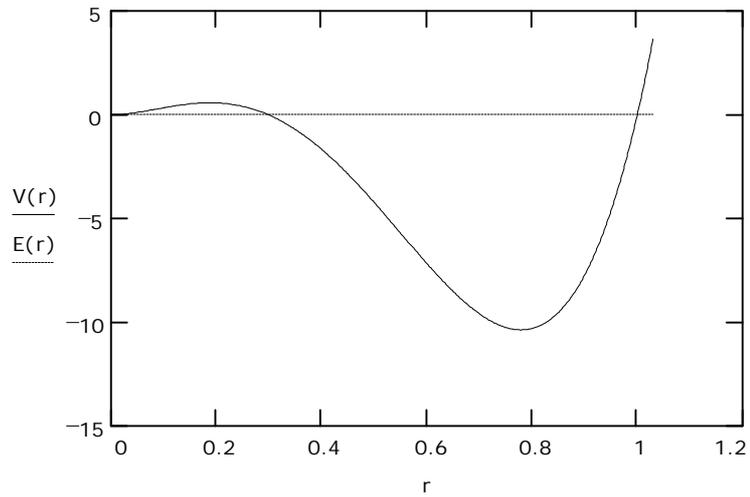

Fig. 1

In the second example, $a = -0.1$ $(a < 0)$. The result appears in Fig. 2.

Here $R_2 = 0.732$ , $R_m = 0.896$ , $R_1 = 0.963$ , $R_M = 1$

and $V(R_2) = 0.222$ , $V(R_1) = -0.154$

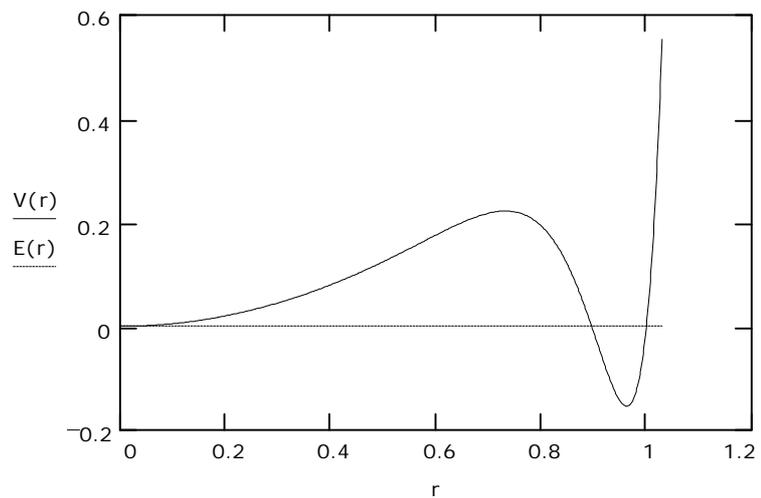

Fig. 2

In the third example, $a = 0.2$ $(0 < a < 1)$. This is plotted in Fig. 3. In this case, there is no maximum because $R_2$ in (3) is complex. Besides, we observe that the limit of $V(0^+)$ is infinite.

Here $\quad R_m = 1 \quad , \quad R_1 = 1.096 \quad , \quad R_M = 1.246$

and $\quad V(R_1) = -0.792$

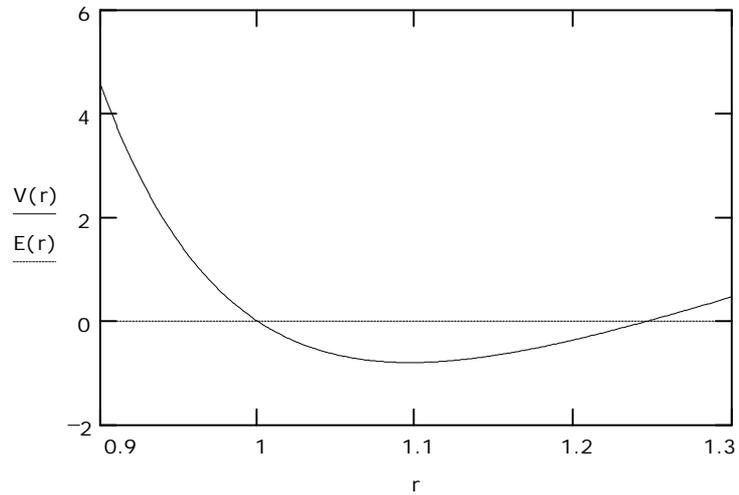

Fig. 3

In the fourth example, $a = 1.6$ $(a > 1)$. The values are (see Fig. 4):

$R_2 = 0.33 \quad , \quad R_m = 1 \quad , \quad R_1 = 3.66 \quad , \quad R_M = 5.8$

and $\quad V(R_2) = 69.644 \quad , \quad V(R_1) = -317.541$

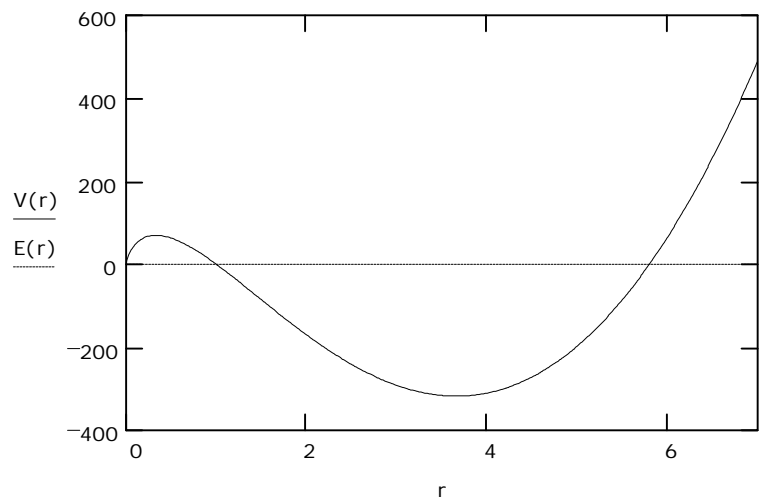

Fig. 4

## III. Harmonic Behaviour

We arrive at the harmonic limit, when we adopt $a = 1$. This conduct is illustrated in Fig. 5. The value of the other parameters is identical to that of the other figures.

The form of the harmonic potential, $V_H(r)$, is

$$V_H(r) = \frac{1}{2} \cdot m \cdot \omega^2 \cdot \left[ (r - r_o)^2 - A_o^2 \right]$$

and the values are:

$$R_2 = \left[ r_o + \frac{-r_o - \sqrt{r_o^2 + 4 \cdot a \cdot (a - 1) \cdot A_o^2}}{2 \cdot a} \right]^a = 0 \quad , \quad R_m = 1$$

$$R_1 = \left[ r_o + \frac{-r_o + \sqrt{r_o^2 + 4 \cdot a \cdot (a - 1) \cdot A_o^2}}{2 \cdot a} \right]^a = r_o = 2 \quad , \quad R_M = 3$$

Here the limit of $V(0^+) = V_H(0) = 150$ and $V(R_1) = -50$

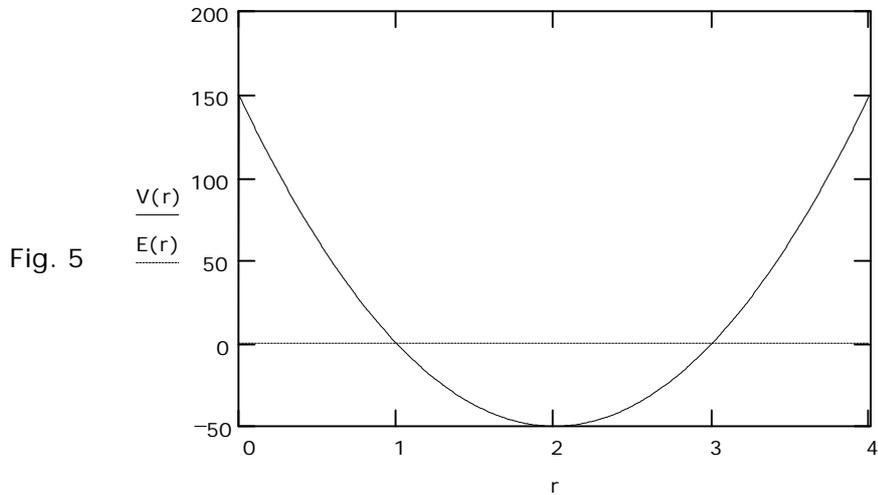

Fig. 5 $\quad \frac{V(r)}{E(r)}$

## IV. Oscillatory Solution

To find the general solution of the oscillatory motion, of energy $E = 0$, in the interval between $R_m$ and $R_M$, we proceed, in the usual way[1], departing from the energy equation:

$$t - t_o = \int_{r(t_o)}^{r(t)} \frac{1}{\sqrt{\frac{2}{m} \cdot (0 - V(r))}} dr = I \qquad (6)$$

The function $I$, can be worked out easily by performing the following change of variable

$$r = \left(r_o + A_o \cdot sen(\omega \cdot s)\right)^a \qquad (7)$$

In terms of $s$, $-I-$ adopts the form:

$$I = \frac{1}{a \cdot \omega} \cdot \int_{s(t_o)}^{s(t)} \frac{a \cdot \omega \cdot A_o \cdot cos(\omega \cdot s)}{\left(r_o + A_o \cdot sen(\omega \cdot s)\right)^{a-a+1} \cdot \frac{A_o \cdot cos(\omega \cdot s)}{r_o + A_o \cdot sen(\omega \cdot s)}} ds =$$

$$= \int_{s(t_o)}^{s(t)} 1 \, ds = s(t) - s(t_o) \qquad (8)$$

Thus, coming back to the r variable and taking for convenience $t_o = 0$, we easily find

$$r = \left(r_o + A_o \cdot sen(\omega \cdot t + \phi_o)\right)^a \qquad (9)$$

The phase, $\phi_o$, is given by:

$$\phi_o = arsen\left(\frac{r(0)^{\frac{1}{a}} - r_o}{A_o}\right) \qquad (10)$$

Adopting $\phi_o = 0$, we have:

$$r(0) = r_o^a \qquad (11)$$

## V. Final Comments

We have presented a new parametric family of anharmonic potentials for which one is able to obtain closed analytical solutions for the trajectories of zero-energy.

The degree of anharmonicity is expressed by the departure of the exponent -a- from the value 1 . Several illustrative examples have been provided.

## *Acknowledgements*